\pdfoutput=1
\documentclass[10pt,twocolumn]{article}

\usepackage{graphicx}
\usepackage{float}
\usepackage[english]{babel}
\usepackage{amsmath}
\usepackage[]{authblk}

\usepackage[utf8]{inputenc}
\usepackage[T1]{fontenc}
\usepackage{verbatim}

\usepackage{color}

\usepackage[hyphens]{url}
\usepackage{pdfpages}

\begin{document}

\title{\LARGE \bfseries News Cohesiveness: an Indicator of Systemic Risk in Financial Markets}

\author[1]{Matija Piškorec}
\author[1]{Nino Antulov-Fantulin}
\author[2]{Petra Kralj Novak}
\author[2]{Igor Mozetič}
\author[2]{Miha Grčar}
\author[3]{Irena Vodenska}
\author[1]{Tomislav Šmuc}
\affil[1]{Laboratory for Information Systems, Division of Electronics, Ruđer Bošković Institute, Croatia}
\affil[2]{Department of Knowledge Technologies, Jožef Stefan Institute, Slovenia}
\affil[3]{Department of Administrative Sciences, Metropolitan College, Boston University, USA}

\maketitle

\begin{abstract}

Motivated by recent financial crises significant research efforts have been put into studying contagion effects and herding behaviour in financial markets. Much less has been said about influence of financial news on financial markets. We propose a novel measure of collective behaviour in financial news on the Web, News Cohesiveness Index (NCI), and show that it can be used as a systemic risk indicator. We evaluate the NCI on financial documents from large Web news sources on a daily basis from October 2011 to July 2013 and analyse the interplay between financial markets and financially related news. We hypothesized that strong cohesion in financial news reflects movements in the financial markets. Cohesiveness is more general and robust measure of systemic risk expressed in news, than measures based on simple occurrences of specific terms. Our results indicate that cohesiveness in the financial news is highly correlated with and driven by volatility on the financial markets.

\end{abstract}

\section*{Introduction}

With the growth of Internet the relationship between online information and financial markets has become a subject of ever increasing interest. Online information offers with respect to its origin and purpose and reflects either interest of some profile of users in the form of query or knowledge about certain topic in the form of news blogs or reports. 

Financial markets are strongly information-driven and these effects can be seen by studying either search query volumes or social media sentiment. Many studies have analysed the effects of search query volumes of specific terms with movements in financial markets of related items~\cite{Ruiz2012}. Bordino et al.~\cite{Bordino_Battiston_2012} show that daily trading volumes of stocks traded in NASDAQ 100 are correlated with daily volumes of Yahoo queries related to the same stocks, and that query volumes can anticipate peaks of trading by one or more days. Dimpfl et al.~\cite{Dimpfl_Jank_2012} report that the Internet search queries for term ``dow'' obtained from Google Trends can help predict Dow Jones realized volatility. Vlastakis et al.~\cite{Vlastakis_2012} study information demand and supply using Google Trends at the company and market level for 30 of the largest stocks traded on NYSE and NASDAQ 100. Chauvet et al.~\cite{Chauvet_fear2012} devise an index of investor distress in the housing market, housing distress index (HDI), also based on Google search query data. Preis et al.~\cite{Moat_Stanley_2013} demonstrate how Google Trends data can be used for designing a market strategy or defining a future orientation index~\cite{preis-forward}. 

In principle, different effects between information sources and financial markets are expected considering news, blogs or even Wikipedia articles~\cite{Moat2013}. Andersen et al.~\cite{Andersen2007} characterize the response of US, German and British stock, bond and foreign exchange markets to real-time U.S. macroeconomic news. Zhang and Sikena exploit~\cite{Zhang_Skiena_2011} blog and news and build a sentiment model using large-scale natural language processing to perform a study on how a company's media frequency, sentiment polarity and subjectivity anticipate or reflect stock trading volumes and financial returns. Chen et al.~\cite{Chen_2012} investigate the role of social media in financial markets, focussing on single-ticker articles published on Seeking Alpha - a popular social-media platform among investors. Mao et al.~\cite{Huina_2011} compare a range of different online sources of information (Twitter feeds, news headlines, and volumes of Google search queries) using sentiment tracking methods and compare their value for financial prediction of market indices such as the DJIA (Dow Jones Industrial Average), trading volumes, and implied market volatility (VIX), as well as gold prices. Casarin and Squazzoni~\cite{Casarin2013} compute Bad News Index as weighted average of negative sentiment words in headlines of three distinct news sources.
 
The idea of cohesiveness of news as a systemic financial risk indicator is related to recent works studying mimicry and co-movement in financial markets as phenomena reflecting systemic risk in financial systems~\cite{debtRank, IrenaRiskProp, Quax2013, Harmon_2011, Kenett2011}. Harmon et al.~\cite{Harmon_2011} show that the last economic crisis and earlier large single-day panics were preceded by extended periods of high levels of market mimicry - direct evidence of uncertainty and nervousness, and of the comparatively weak influence of external news. Kennet et al.~\cite{Kenett2011} define an index representing the balance between the stock correlations and the partial correlations after subtraction of the index contribution and study the dynamics of S\&P 500 over the period of 10 years (1999-2010).

The idea of cohesiveness as a measure of news importance is simple: if many sources report about same events then this should reflect their importance and correlate with the main trends in financial markets. However, in order to capture the trends of systemic importance one must be able to track different topics over majority of relevant online news sources. In other words, one needs: (i) an access to the relevant news sources and (ii) a comprehensive vocabulary of terms relevant for the domain of interest. We satisfy the second prerequisite for systemic approach through the use of large vocabulary of financial terms corresponding to companies, financial institutions, financial instruments and financial glossary terms. To satisfy the first prerequisite, in our analysis we rely on financial news documents extracted by a novel text-stream processing pipeline NewStream~\cite{KraljNovak2014} (\texttt{http://newstream.ijs.si/}), from a large number of Web sources. These texts are then filtered and transformed into a form convenient for computing NCI for the particular period of time. 

We show that importance of financial news can be measured in a more systemic way than via sentiment towards individual entities or number of occurrences of individual terms, and that strong cohesiveness in news reflects the trends in financial markets. There is already a strong evidence linking co-movement of financial instruments to systemic risk in financial markets~\cite{Kenett2011}. We hypothesize that cohesiveness of financial news reflects in some part this systemic risk. Our News Cohesiveness Index (NCI) captures average mutual similarity between the documents and entities in the financial corpus. If we represent documents as sets of entitites then there are two alternative views on similarity: (i) two documents are more similar than some other two documents if they share more entities and (ii) two entities are more similar than some other two entities if they co-occur in more documents. We construct NCI so that the overall similarity in a corpus of documents is equal regardless on the view we choose to adopt. 

We analyse the NCI in the context of different financial indices, their volatility, trading volumes, as well as Google search query volumes.
We show that NCI is highly correlated with volatility of main US and EU stock market indices, in particular their historical volatility and VIX (implied volatility of S\&P500). Furthermore, we demonstrate that there is a substantial difference between aggregate term occurrence and cohesiveness in their relations toward financial indices.

\section*{Results}

\subsection*{News Cohesiveness Index}
In order to measure the herding effects in financial news we introduce a News Cohesiveness Index (NCI) - a systemic indicator that quantifies cohesion in a collection of financial documents. A starting point for the calculation of NCI is a \emph{document-entity matrix} that quantifies occurrences of entities in each individual document collected over certain period of time. We use concept of entity (instead of e.g. term) to represent different lexical appearances of some concept in texts. In our case we use a vocabulary of entities that includes financial glossary terms, financial institutions, companies and financial instruments. The full taxonomy of entities is available in Supplementary Information section 3. We start with the definition of occurrence, which says whether some entity is present or not in some document, regardless of how many times it occurs in the document. This makes document-entity matrix $A$ a binary matrix:

\begin{equation}
A_{i,j} = \left\{
\begin{array}{l l}
1 & \textrm{if entity $e_j$ is in document $d_i$} \\
0 & \textrm{otherwise}.
\end{array} \right.
\end{equation}

$A$ is an $m \times n$ matrix, where $m$ is the number of documents published in selected time period and $n$ is the total number of entities we monitor. Document-entity matrix $A$ also corresponds to a biadjacency matrix of a bipartite graph between documents and entities. An edge between document $d_i$ and entity $e_j$ exists if the entity $e_j$ appears in the document $d_i$. 

The overall similarity in the collection of documents should be equal regardless whether we choose to view it as the similarity between the \emph{documents} or between the \emph{entities}. To achieve this we define the similarity as the \emph{scalar product} of either document pairs $\langle d_i, d_j \rangle$ or entity pairs $\langle e_i, e_j \rangle$. Now we define NCI as a Frobenius norm of scalar similarity matrix between all pairs of documents $C_{ij}^d = \langle d_i, d_j \rangle$ or entities $C_{ij}^e = \langle e_i, e_j \rangle$:
 
\begin{equation}
NCI = \parallel C \parallel_F = \sqrt{ \sum^{m}_{i=1} \sum^{m}_{j=1} \| C_{ij} \|^2 },
\end{equation}

where $C$ is either $C^d$ or $C^e$. Frobenius norms of both document-document similarity matrix $C^d = AA^T$ and entity-entity similarity matrix $C^e = A^TA$ are equal, therefore the cohesion is conserved whether we measure it as the \emph{document} or the \emph{entity} similarity:

\begin{equation}
\parallel C^d \parallel_F = \parallel AA^T \parallel_F =  \parallel A^TA \parallel_F = \parallel C^e \parallel_F.
\label{equ:conservation-cohesiveness}
\end{equation}

\begin{figure}
\centering
\includegraphics[width=\columnwidth]{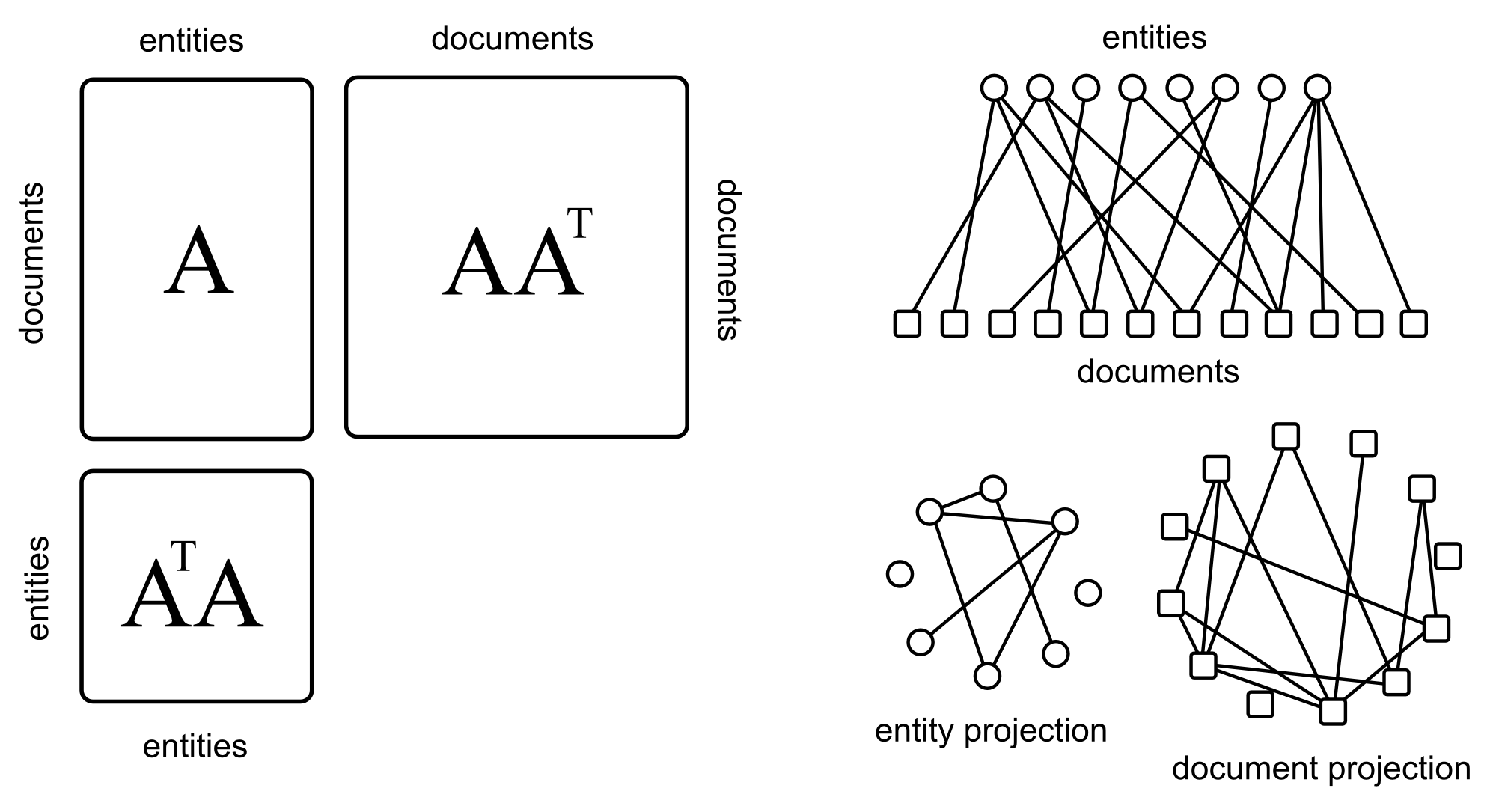}
\caption{\textbf{Matrix and network representations of document-entity matrix.} Matrix representation of document-document and entity-entity similarity matrices (left), and the corresponding network representations of entity and document projections (right). Frobenius norms of the two similarity matrices correspond to the sum of squares of connection weights in two projections, and they are equal, which means that cohesiveness is conserved in both projections.}
\label{fig:projection-matrix-network-representation}
\end{figure}

In the network representation these two similarity matrices correspond to two projections of a bipartite graph of the original document-entity matrix, as illustrated in Figure~\ref{fig:projection-matrix-network-representation}. Moreover, one can exploit properties of the Frobenius norm of scalar similarity matrix and express cohesiveness as a function of singular values of document-entity matrix $A$ (proof in the Supplementary Information section 2): 

\begin{equation}
NCI = \sqrt{ \sum^{k}_{i=1} \sigma_{i}^4 },
\label{equ:singular-values}
\end{equation}

where $\sigma_{i}$ are the $k$ largest singular values of matrix $A$ in a singular value decomposition:
\[
A = U \times S \times V^T.
\]
Because singular values are calculated on the original document-entity matrix and not its document or entity projections, we claim that we capture an \emph{intrinsic} property of the corresponding document-entity matrix that is invariant to projection. This can also be inferred from the fact that the eigenvalues of similarity matrices $AA^{T}$ and $A^{T}A$ are equal and they correspond to the singular values of document-entity matrix $A$. 

This approach can be beneficial for large document-entity matrices as it is much more efficient in terms of time and memory compared to explicit calculation of similarity matrix. We can calculate just first $k$ values incrementally, until we reach the desired accuracy of NCI (see Supplementary Information section 1). In practice, only a small number of singular values is enough to calculate NCI up to the desired precision.

As the number of documents is changing each day while the number of entities stays constant, all NCI indices in our analyses are normalized with the number documents $1/m$ in the corpus. We have statistically confirmed that the NCI is largely above the level of fluctuations of cohesiveness random null model (see Supplementary Information section 2).

\subsubsection*{Semantic partitions of NCI}
\label{sec:semantic-partitioning}

\begin{figure}
\centering
\includegraphics[width=\columnwidth]{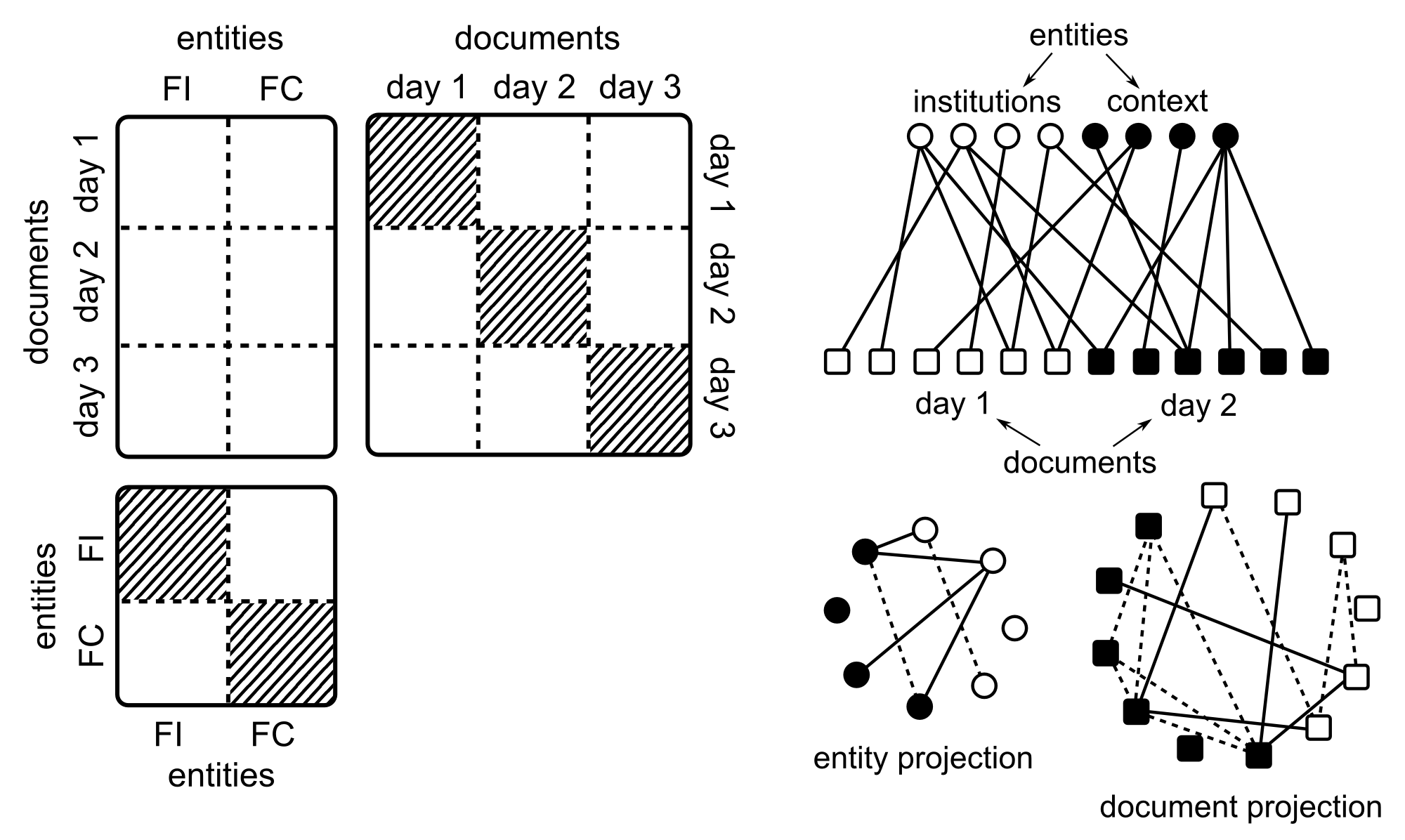}
\caption{\textbf{Semantic partitioning.} Semantic partitioning for two entity semantic groups - ``Financial Institutions'' and ``Financial Context'', and three document semantic groups - ``day 1'', ``day 2'' and ``day 3''. Frobenius norm in shaded regions quantifies cohesiveness within each semantic group, while Frobenius norms of all other regions quantify cohesiveness based on pairs of semantic groups.}
\label{fig:semantic-partitioning}
\end{figure}

Sometimes it is interesting to perform detailed analysis of which groups of entities or documents contribute the most to the overall cohesiveness. For this purpose we can divide entities or documents into groups using any appropriate semantic criteria and calculate cohesiveness for each group separately or between pairs of groups. Semantic partitions in the entity projection are created via grouping of entities in mutually disjoint groups, defined by their taxonomy labels (hence semantic interpretation). On the other hand, semantic partitions in the document projection can be created via grouping of entities either by their temporal or source membership. Figure~\ref{fig:semantic-partitioning} illustrates the concept of partitioning in the context of different projections.

\begin{figure*}
\centering
\includegraphics[width=0.9\textwidth]{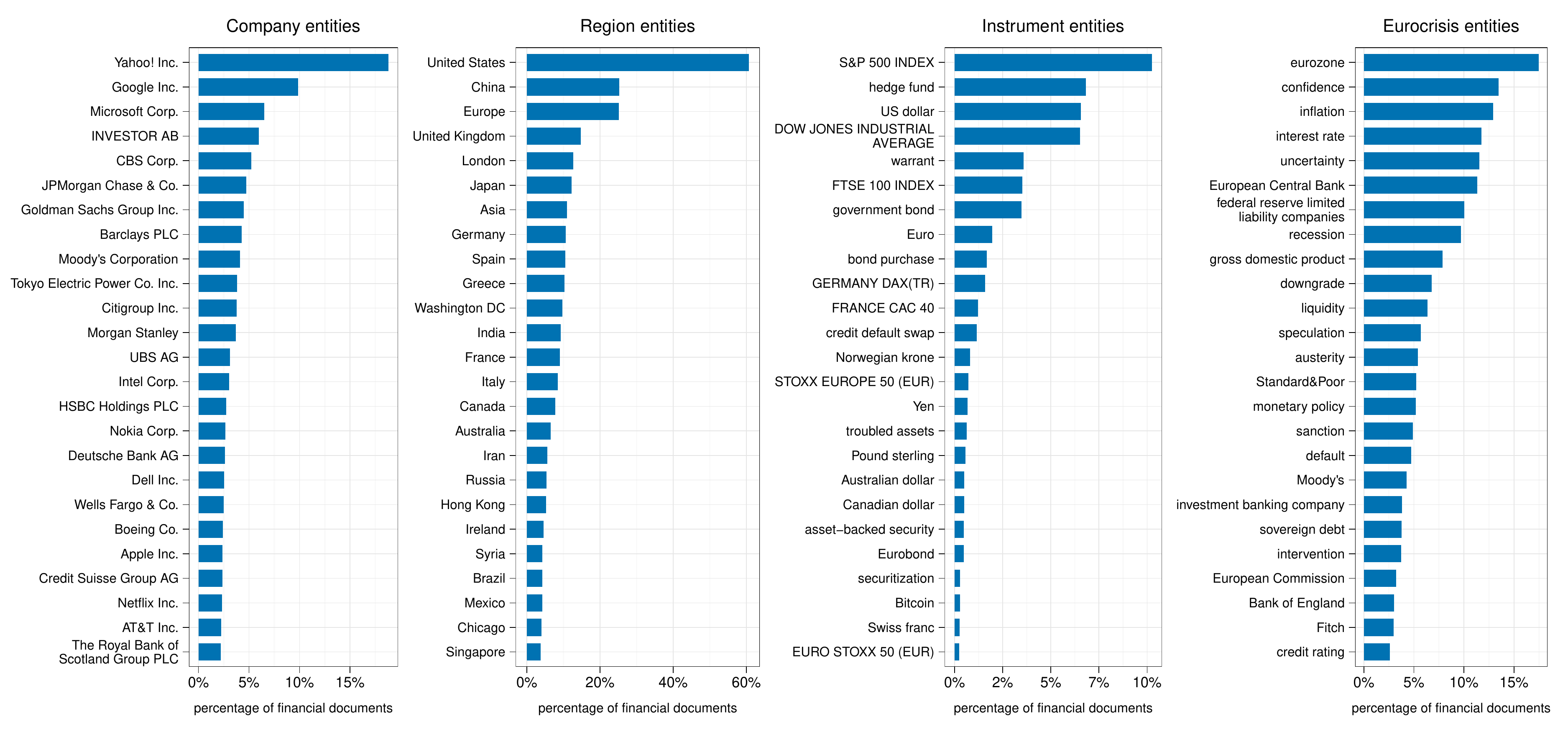}
\caption{\textbf{Occurrences of most frequent entities in each of the semantic partitions.} One of the most frequent entities are those defining major regions corresponding to the world's leading financial markets: United States, China, Europe, United Kingdom, London, Japan, Germany. Considering the frequency of United States, it is no surprise that majority of other frequent entities, from companies to instruments, are also tied to the US financial market and terminology.}
\label{fig:semantic_components_occurrences}
\end{figure*}

We can calculate cohesiveness separately for each semantic group or a combination of semantic groups. Note that even in this case we do not need to explicitly calculate similarity matrices (see Supplementary Information section 1). Following the taxonomy of entities described in Supplementary Information section 3 we defined four semantic groups: companies, regions, financial instrument and Euro crisis terms. Figure~\ref{fig:semantic_components_occurrences} shows the most frequent entities in each of the semantic partitions, based on the news corpus collected over the period of analysis.

\subsection*{NCI in relation to financial markets and query volumes}

In order to asses NCI's utility as a systemic risk indicator, we use correlations analysis and Granger causality tests against the pool of financial market and information indicators. The analysis should also provide deeper insight into the interplay between news, trends in financial markets and behaviour of investors. We adopt terminology from ~\cite{Vlastakis_2012}, and treat our news based indicators (NCI variants and entity occurence) as indicators of the information supply in online media, while volumes of Google Search Queries will be treated as indicators of information demand or as a proxy of investor interest.

We group indicators as follows:
\begin{itemize}
      \item \textbf{Inormation supply indicators:} - cohesiveness index based on all the news (NCI) from NewStream, cohesiveness index based only on filtered financial news (NCI-financial) from NewStream, total entity occurrences based on the aggregate from all news documents, and total entity occurrences based on strictly financial documents of NewStream. 
      \item \textbf{Information demand indicators:} - these are volumes of Google Search Queries (GSQ) for 4 finance/economy related categories from Google Finance (from Google Domestic trends - Finance\&Investment, Bankruptcy, Financial Planning, Business).
      \item \textbf{Financial market indicators:} - these include daily realized volatilities, historical volatilities and trading volumes of major stock market indices (S\&P 500, DAX, FTSE, Nikkei 225, Hang Seng) as well as implied volatilities of S\&P500 (VIX). 
\end{itemize}

Details on preparation of individual indicators are given in Methods section.

\begin{figure*}
\centering
\includegraphics[width=0.8\textwidth]{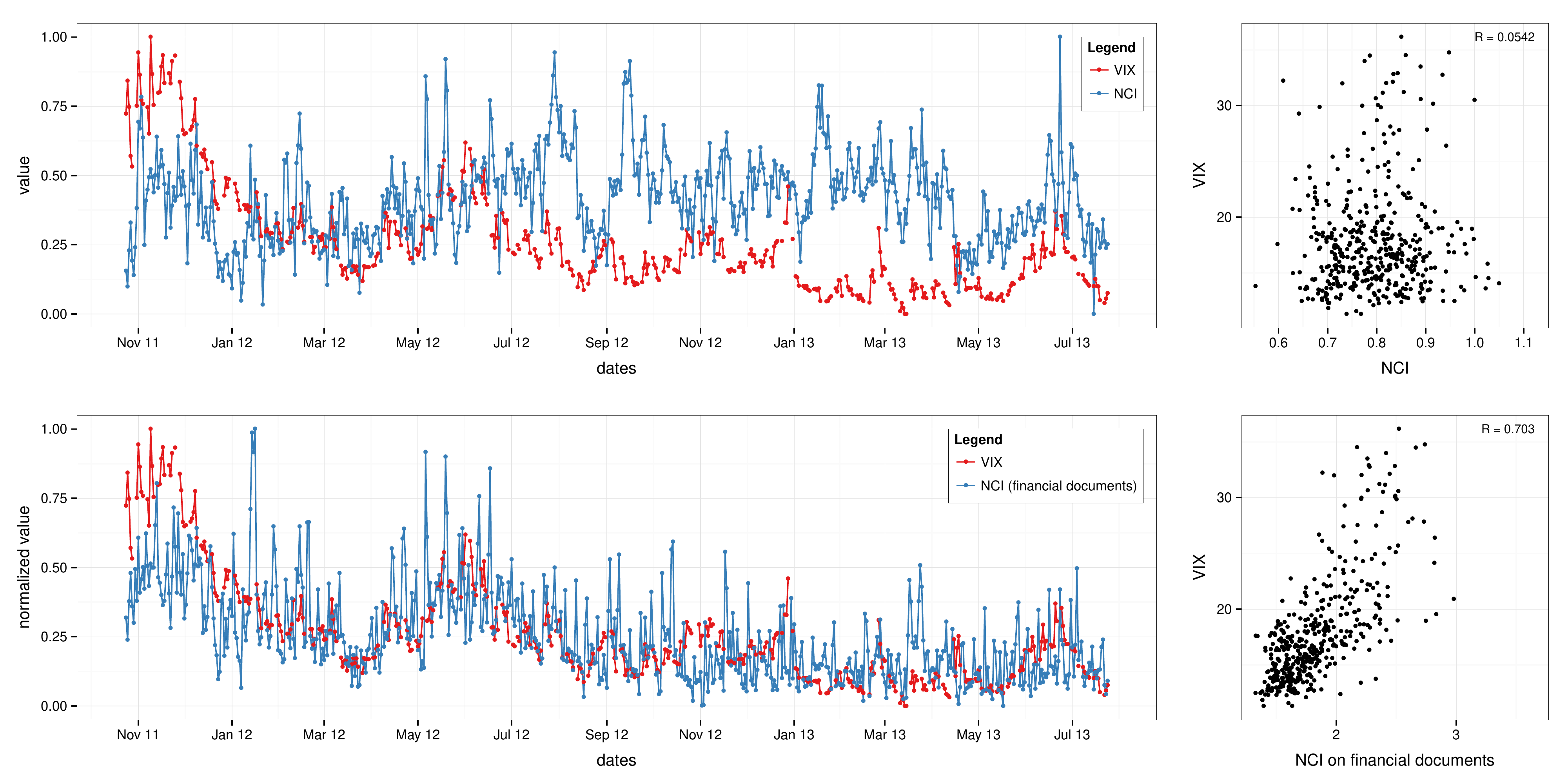}
\caption{\textbf{Comparison of NCI and VIX time series.} NCI - calculated on all news (top), NCI-financial - calculated only on strictly financial news (bottom) and their correlation to VIX (right). Time series for NCI cover 640 days from 24th October 2011 to 24th July 2013. Time series for VIX cover 439 working days in the same period.}
\label{fig:nci_vs_indices}
\end{figure*}

We start the analysis with a simple comparison of NCI calculated on all news and NCI calculated on filtered financial news. Figure~\ref{fig:nci_vs_indices} shows dynamics of NCI, and NCI-financial in comparison to VIX (implied volatility of S\&P 500, the so called ``fear factor''~\cite{Vodenska2013}). Scatter plots on the right show that correlation of VIX and NCI-financial is significantly higher than VIX and NCI. This is a first illustration of the importance of the filtering the right content for the construction of indicators from texts. For more details on how filtering affects correlations with other indices see Supplementary Information section 3.

\begin{figure*}
\centering
\includegraphics[width=\textwidth]{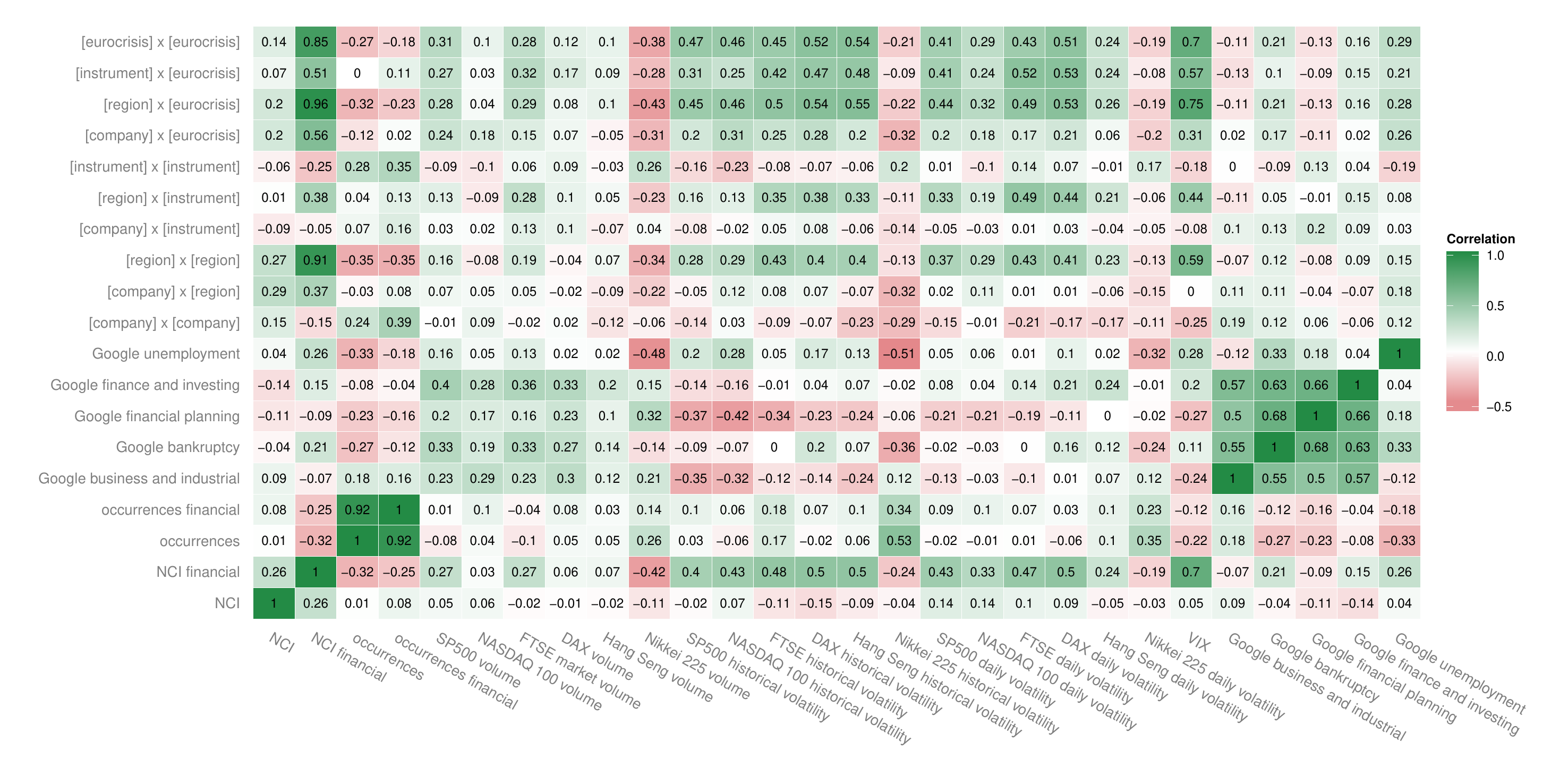}
\caption{\textbf{Pearson correlation matrix between all indices.} Indices include: NCI on all documents; NCI-financial (calculated on selected financial documents) and its semantic components; entity occurrences, implied volatility of S\&P 500 (VIX), realized historical, and daily volatilities of main stock market indicators (S\&P 500, NASDAQ 100, FTSE, DAX, Nikkei, Hang Seng) and Google Search Query indicators - Business and industrial, Bankruptcy, Financial Planning, Finance and investing, Unemployment. Corresponding p-values for all the correlations are given in Supplementary Information section 5.}
\label{fig:corr_matrix}
\end{figure*}

Figure~\ref{fig:corr_matrix} shows Pearson correlation coefficients between different information indicators and financial market indicators. Corresponding p-values are calculated using a permutation test and are available in Supplementary Information section 5. All correlations reported in this article have p-value $<10^{-4}$ unless explicitly stated.

Interesingly, the correlations between total entity occurrences, NCI and NCI-financial are relatively low, confirming that cohesiveness captures very different signal from the entity occurrences. Furthermore, correlations between total entity occurrences, NCI and financial indices are, on average, much lower than correlations between NCI-financial and financial indices. Relatively low correlation between NCI-financial and NCI confirms importance of filtering out strictly financial market-related articles from the NewStream, rather than having all the articles that contain some of the entities from the vocabulary. We have performed a more detailed analysis of these effects by studying in parallel behavior of different variants of entity occurrences and NCI-financial using different subsets of the vocabulary and the document space, independently. The main insight gained was that entity occurences become more informative when a smaller vocabulary of the most frequent entities is used, but this requires use of the whole document space. NCI has proven to be much more robust to the choice of both vocabulary and document space (details in Supplementary Information section 6).

Interestingly, the NCI-financial index is highly correlated with implied volatility ($R>0.7$, Figure~\ref{fig:corr_matrix}), as well as with historical and daily realized volatilities ($R>0.4$, Figure~\ref{fig:corr_matrix}). These correlations are much higher than the correlations of the GSQ categories ($R<0.3$, Figure~\ref{fig:corr_matrix}). In contrast to NCI-financial, GSQ categories exhibit relatively stronger correlations with stock trading volumes ($0.3<R<0.4$, Figure~\ref{fig:corr_matrix}). Google Bankruptcy and Google Unemployment are significantly correlated with NCI-financial (correlation above 0.2, Figure~\ref{fig:corr_matrix}), which is most probably due to similarities in vocabulary  used in constructing NCI-financial and respective GSQ indicators.

A more in depth picture of the news cohesiveness index is obtained when observing individual semantic components of NCI-financial and their correlation patterns with financial and Google search query indicators. Semantic components based on \texttt{[region]} and \texttt{[eurocrisis]} taxonomy categories all have similar correlation patterns to NCI-financial (with correlation above 0.7 for \texttt{[eurocrisis]} and above 0.5 for \texttt{[region]}, Figure~\ref{fig:corr_matrix}); this also shows that these categories are most important for the behavior of NCI-financial. On the other hand, semantic components based on \texttt{[company]} and \texttt{[instrument]} exhibit quite different, in many parts, opposite correlation patterns (with correlations close to 0 or even negative). It is interesting to note that both the NCI-financial and GSQ indicators have strong negative correlation with Nikkei 225 volatility and trading volume (up to -0.4 for NCI-financial and up to -0.5 for GSQ-unemployment). 

\subsection*{Granger causality relations}

The Granger-Causality test (GC test) is frequently used to determine whether a time series $Y(t)$ is useful in forecasting another time series $X(t)$. The idea of the GC test is to evaluate if $X(t)$ can be better predicted using both the histories of $X(t)$ and $Y(t)$ rather than using only the history of $X(t)$ (i.e. $Y(t)$ Granger-causes $X(t)$). The test is performed by regressing $X(t)$ on its own time-lagged values and on those of $Y(t)$ included. An F-test is used in examining if the null hypothesis that $X(t)$ is not Granger-caused by $Y(t)$ can be rejected.

\begin{figure*}
\centering
\includegraphics[width=0.95\textwidth]{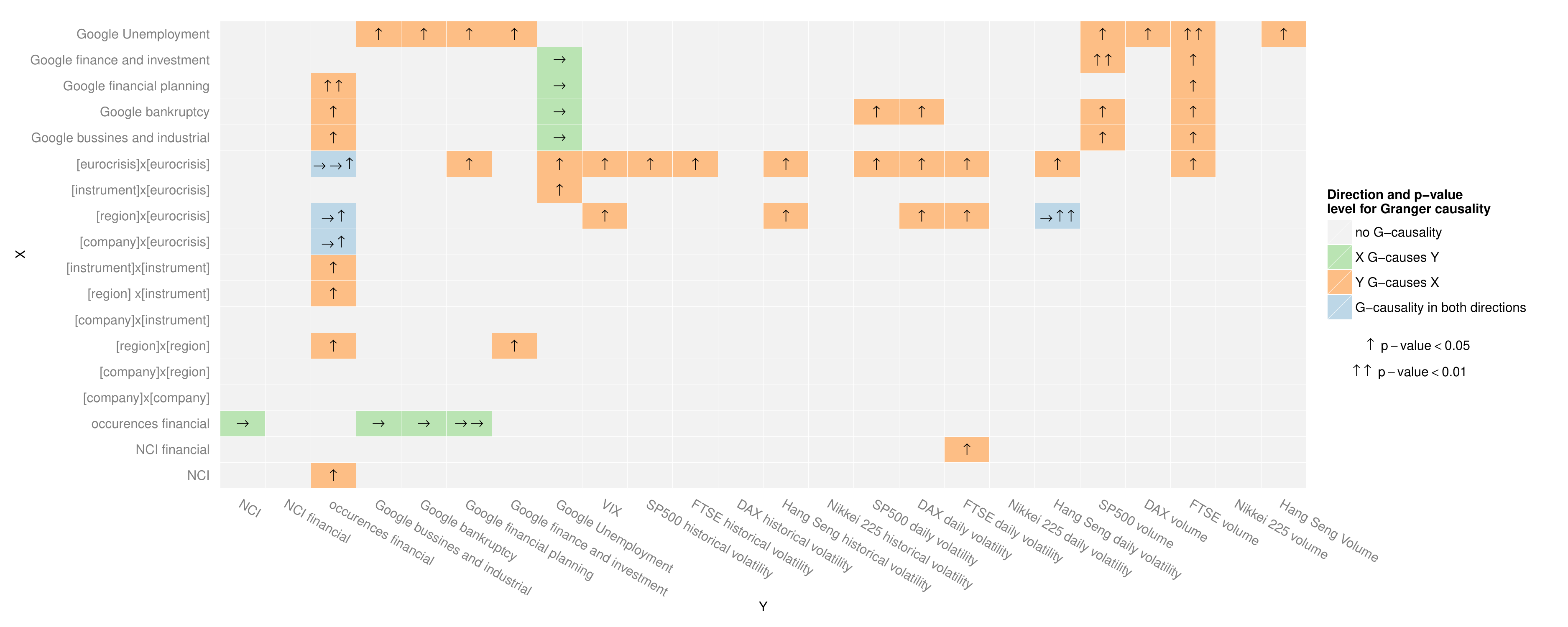}
\caption{\textbf{Granger causality tests.} Results of Granger causality tests for the mutual impacts between information and financial indicators. Colors indicate the direction of G-causality ($X \rightarrow Y$ or $Y \rightarrow X$) as well as bidirectional G-causality. ($X \leftrightarrow Y$) at two level of significance (F-test p-value < 0.01 and p-value < 0.05).}
\label{fig:table_GC_bidirect}
\end{figure*}

In Table \ref{fig:table_GC_bidirect} we show results of pairwise G-causality tests between information supply and demand indicators and financial indicators. Cells of the table give both directionality ($X \rightarrow Y$, $Y \rightarrow X$ or bidirectional $X \leftrightarrow Y$) and significance at two levels of F-test (p-values $\leq0.01$; $\leq0.05$).
Besides GC testing NCI-financial and its semantic components at higher taxonomy levels, we show also results obtained for NCI (non-filtered news NCI) and total entity occurences as a baseline.  

The results in Table~\ref{fig:table_GC_bidirect} paint a much different picture than the correlation study. Firstly, Granger causality seems to be almost exclusively directed from financial to information world, with single bidirectional exception between \texttt{[region]x[eurocrisis]} semantic component of NCI-financial and Hang Seng daily realized volatility.

Our financial news indicator NCI-financial seems to be G-caused solely by FTSE daily volatility. This finding is in contrast with the fact that the NCI-financial is strongly correlated with several other indicators like implied volatility (VIX) ($R>0.7$, Table~\ref{fig:corr_matrix}). However, two of the semantic components \texttt{ [eurocrisis]x[eurocrisis]} and \texttt{[region]x[eurocrisis]} are strongly G-caused by implied volatility, historical and daily volatilities of most of the  major stock market indices. On the other hand, the GSQ categories seem to be mostly GC driven by trading volumes, almost exclusively of US and UK financial market (S\&P 500 and FTSE).

GSQ indicators seem to be divided in two groups by their G-causality: (i) those that are G-caused mainly by trading volumes (Business and industrial, Bankruptcy, Financial planning and Finance and investment) and total entity occurrences in the news, and (ii) those that are strongly G-caused by all other GSQ categories (Unemployment). Interestingly, total entity occurrence in the news, seem to be the strongest G-causality driver of the GSQ volumes, while two of the semantic components of NCI-financial are G-caused by GSQ Finance and investment and Financial planning.

\section*{Discussion}

This work introduces a new indicator of financial news importance based on a concept of cohesiveness of texts, from large corpora of news and blogs sources. In contrast to indicators introduced by other authors which are based on sentiment modelling, NCI measures cohesiveness in the news by approximating the average similarity between texts. 

Our correlation results confirm the main hypothesis that cohesiveness of the financial news is a signal that is strongly correlated with systemic financial market indicators in particular volatilities of major stock exchanges. 

The analysis of Granger causality tests over a pool of financial and information related indicators suggests that NCI-financial is mainly related with the volatility of the market. In our analysis most important semantic components of NCI-financial are mainly G-caused by implied (VIX), historical and daily volatilities. This implies effects from both short term and long term risks in the financial market. The only exception (bidirectional causality between \textit{[region]x[eurocrisis]} and Hang Seng daily volatility) might be plausibly explained as a time zone effect. This does not seem to be the case for GSQ indicators  which are mainly driven by trading volumes, with the exception of GSQ Unemployment, which seems to be driven mostly by search volumes of other GSQ categories. Similar to findings of some previous studies~\cite{Da_2011,Casarin2013}, in which aggregate sentiment or financial headline occurrence were used as measures of state of the financial market, information supply co-movement as measured by NCI-financial, seem to be primarily caused by trends on the financial market rather than the opposite. We find that similar results holds also for the GSQ categories which approximate information demand side in our case. G-causality patterns show, similarly to correlation, that cohesiveness captures quite different signal with respect to total  entity occurrence; the results also suggest the presence of somewhat circular interplay between information supply and information demand indicators. For example, total entity occurrence is G-causing three of the GSQ categories (Business and industry, Bankruptcy and Financial planning), while Financial planning and Unemployment are G-causal for semantic components \texttt{[instrument]x[eurocrisis]} and \texttt{[eurocrisis]x[eurocrisis]}, which suggests feedback mechanisms between news and search behaviour. 

In comparison with the findings of studies which used simpler measures of news importance or sentiment, we find that financial news cohesiveness reflects the level of the volatility in the market and is GC driven both by current level of volatility and implied volatility, while GSQ volumes are driven mainly by trading volumes. Impact of news cohesiveness and GSQ volumes in the reverse direction, as determined by GC tests, is only weakly implied in case of semantic components of NCI-financial and Hang Seng index. This is not in line with previous works that report predictive utility (mostly for GSQ volumes) with respect to certain financial instruments. However, one has to bear in mind that the results of GC tests reflect average of lagged correlations between indicators over the specific period in time (in our case Oct 2011 - Jul 2013). It is possible that direction of causality between information and financial indicators changes in time, but this was hard to detect in our data due to the limited length of time series. Another possible reason for different results is that most of previous works were based on a limited number of Google search query terms, typicaly more closely related to the particular stock market index of interest. In principle, this is different from volumes of GSQ term categories in our case, which reflect aggregates over larger number of different query terms. GSQ categories closely resemble the concept of semantic components and it is possible that the application of the  concept of cohesiveness, if adapted to GSQ category volumes, may produce signals more predictive with respect to financial market trends.

\section*{Methods}

\subsection*{Data}
Access to structured information about the financial markets with its various instruments and indicators is available for several decades, but systematic quantification of unstructured information hidden in news from diverse Web sources is of relatively recent origin. 

We base our analyses on a newly created text processing pipeline - NewStream~\cite{KraljNovak2014}, designed and implemented within the scope of EU FP7 projects FIRST (\texttt{http://project-first.eu/}) and FOC (\texttt{http://www.focproject.eu/}). NewStream continuously downloads articles from more than 200 worldwide news sources, extracts the content and stores complete texts of articles. It is a domain independent data acquisition pipeline, but biased towards finance by the selection of news sources and the taxonomy of entities relevant for finance.
 
For the purpose of filtering, efficient storing and analytics, expert based financial taxonomy and vocabulary of entities and terms have been created, containing names of relevant financial institutions, companies, finance and economics specific terms, etc. The NewStream pipeline has been collecting data since October 2011. In our analyses we use text corpora from October 2011 to June 2013 and we have filtered over 1,400,000 financially related texts stored in the form of document-entity matrices. Full structure of the taxonomy is in the Supplementary Information section 3, and the list of the domains from which most documents were downloaded in the Supplementary Information section 4.

\paragraph*{Filtering of financial documents}
NewStream pipeline downloads articles from more than 200 Web sites of online news and blogs. Moreover, despite the selection of financial news sites, there are many articles which are only indirectly related to finance, such as politics or even sport. To obtain a clean collection of strictly financial texts, we have developed a rule-based model utilizing taxonomy categories as features to describe documents, and a gold standard of human labelled documents ($\approx$ 3500 documents). A machine learned rule-based model is used as a filter for extracting strictly finance related texts from a corpus. This model has a recall of over 50\%, with precision of well over 80\%. The rule-based model for filtering financial documents can be found in Supplementary Information section 3. 

\paragraph*{Financial indicators}
We analyse NCI in comparison to the financial market indicators of worldwide markets and Google Search Query volumes. For that purpose we have downloaded stock market indices from Yahoo finance Web service (\texttt{http://finance.yahoo.com/}): High, Low, Open, Close prices and volume of S\&P 500, DAX, FTSE, Nikkei 225 and Hang Seng index. We also use implied volatility of  S\&P500 (VIX). Implied volatily is calculated for the next 30 days by Chicago Board Options Exchange (CBOE, \texttt{http://www.cboe.com/}) using current prices of indices options. Historical (realized) volatilities are calculated from the past prices of the indices themselves. We use daily prices of individual indices to calculate a proxy of daily realized volatility.

\begin{equation}
\text{daily volatility} = \frac{High_{t}-Low_{t}}{0.5\left(Close_{t}+Close_{t-1}\right)}.
\end{equation}

Historical (realized) volatilities are calculated as standard deviations of daily log returns in the appropriate time window:

\begin{equation}
\text{Historical volatility} = \sqrt{\frac{1}{n}\sum_{t}^{window}\left(\log\left(\frac{p_{t}}{p_{t-1}}\right)\right)^{2}},
\end{equation}

where $p_{t}$ are daily prices, and $n$ is time window. In our analyses we used a window of 21 working days.

\paragraph*{Google Search Query Volumes}
Almost all previous studies used search query volumes of specific terms. Instead, we used Google search query volumes of predefined term categories from Google finance website. We have chosen five categories from Google Domestic Trends that are related to financial market: Business and industrial, Bankruptcy, Financial Planning, Finance and investing, Unemployment. We downloaded YOY (Year Over Year) change values for these categories from Google finance Web service (\texttt{https://www.google.com/finance}).

\subsection*{Granger causality testing}
We have used functions of the R packages \textit{tseries}, \textit{lmtest}, \textit{vars}, \textit{urca} to download and calculate indices, construct joint time series dataset, determine correlations and study Granger causality relations. We have followed the methodology of Toda and Yamamoto~\cite{Toda_1995} for Granger causality testing of non-stationary series. Details of the procedure are given in Supplementary Information section 5.

% % \bibliographystyle{plain}
% % \bibliographystyle{acm}
% \bibliographystyle{naturemag}

% \bibliography{NCIbibliography}

% We include bibliography explicitaly. After running bibtex copy the contents from generated bbl file.

\section*{Acknowledgements}
This work was supported in part by the European commission under the FP7 projects FOC (Forecasting Financial Crises, Measurements, Models and Predictions, grant no. 255987 and FOC INCO, grant no. 297149) and by the Croatian Ministry of Science, Education and Sport project ``Machine Learning Algorithms and Applications''. We would like to thank the following people for helpful discussions: Stefano Battiston, Vinko Zlatić, Guido Caldarelli, Michelangelo Puliga, Tomislav Lipić and Matej Mihelčić.

\section*{Author contributions}
All authors contributed to the writing and editing of the manuscript. MP, NAF and TS carried out the modelling and the analyses. PKN, IM and MG were involved in gathering and processing of the data. IV and TS were involved in interpretation of the results.

\section*{Additional information}

\textbf{Supplementary Information} accompanies this paper. \\
\textbf{Competing financial interests:} The authors declare no competing financial interests. \\
\textbf{Dataset availability:} All data and codes that we used in our analysis is freely available on \texttt{http://lis.irb.hr/foc/data/data.html}.

\section*{Corresponding author}
Correspondence to Tomislav Šmuc, \\email: \texttt{tomislav.smuc@irb.hr}

\includepdf[pages=-]{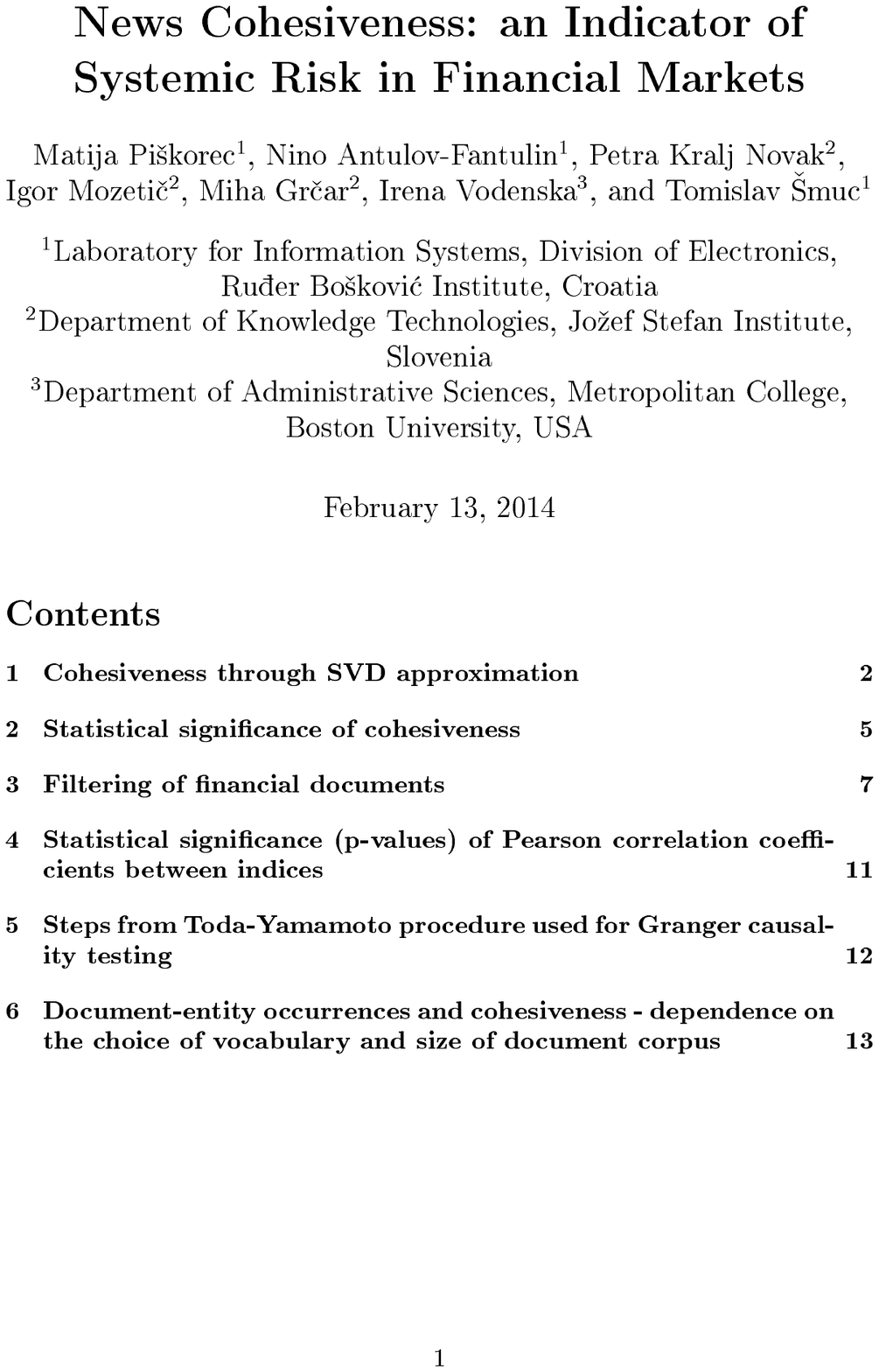}

\end{document}